\documentclass[twocolumn,showpacs,superscriptaddress,prb]{revtex4}
\usepackage{amsbsy,amssymb,amsmath,bm} \usepackage{graphicx}
\newcommand{\bS}{\mathbf{S}} 
\newcommand{\bH}{\mathbf{H}}
 \newcommand{\br}{\mathbf{r}}
\newcommand{\cG}{\mathcal{G}} 
\newcommand{\cE}{\mathcal{E}}
\newcommand{\cP}{\mathcal{P}} \newcommand{\cF}{\mathcal{F}}
\newcommand{\cL}{\mathcal{L}} \newcommand{\cM}{\mathcal{M}}
\newcommand{\ua}{\uparrow} \newcommand{\da}{\downarrow}

\begin{document} \title{Point contact spectroscopy of hopping
transport: effects of a magnetic field}
\author{V. I. Kozub}
\affiliation{A. F. Ioffe  Physico-Technical Institute of Russian
Academy of Sciences, 194021 St. Petersburg, Russia}
\affiliation{Argonne National Laboratory, 9700 S. Cass Av.,
Argonne, IL 60439, USA}
\author{A. A. Zyuzin}
\affiliation{A. F. Ioffe Physico-Technical Institute of Russian
Academy of Sciences, 194021 St. Petersburg, Russia}
\author{O. Entin-Wohlman}
\author{A. Aharony}
\affiliation{Department of Physics and the Ilse Katz Center for
Meso- and Nano-Scale Science and Technology, Ben Gurion
University, Beer Sheva 84105, Israel}
\affiliation{Argonne
National Laboratory, 9700 S. Cass Av., Argonne, IL 60439, USA}
\author{Y. M. Galperin}
\affiliation{Department of Physics and Center for Advanced
Materials and Nanotechnology, University of Oslo, PO Box 1048
Blindern, 0316 Oslo, Norway}
\affiliation{A. F. Ioffe
Physico-Technical Institute of Russian
  Academy of Sciences, 194021 St. Petersburg, Russia}
\affiliation{Argonne National Laboratory, 9700 S. Cass Av.,
Argonne,
  IL 60439, USA}
\author{V. Vinokur}
\affiliation{Argonne National Laboratory, 9700 S. Cass Av.,
Argonne, IL 60439, USA}
\date{\today}
\begin{abstract}

The conductance of a point contact between two hopping insulators is
expected to be dominated by the individual localized states in its
vicinity. Here we study the additional effects due to an external
magnetic field. Combined with the measured conductance, the measured
magnetoresistance provides detailed information on these states
(e.g. their localization length, the energy difference and the hopping
distance between them). We also calculate the statistics of this
magnetoresistance, which can be collected by changing the gate voltage
in a single device. Since the conductance is dominated by the quantum
interference of particular mesoscopic structures near the point
contact, it is predicted to exhibit Aharonov-Bohm oscillations, which
yield information on the geometry of these structures. These
oscillations also depend on local spin accumulation and correlations,
which can be modified by the external field.  Finally, we also
estimate the mesoscopic Hall voltage due to these structures.

\end{abstract} 
\maketitle

% Body of the text
%

\section{Introduction}

\begin{figure}[h] \centerline{
\includegraphics[width=5cm]{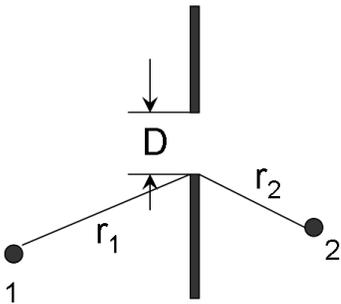} } \caption{Tunnelling paths near
the point contact. \label{fig:fig1}}
\end{figure}

Studies of current-voltage curves of tiny contacts between two
electrodes -- point contact spectroscopy -- are a powerful tool
for the investigation of charge transport through various systems.
It has provided numerous important results for normal metals and
superconductors.\cite{Yanson}  Recently, two of us \cite{KZ} have
studied such point contacts between two {\it hopping insulators}
(HI). When the diameter of the contact, $D$, is smaller than the
typical hopping distance in the bulk, $r_h$, it has been found
that the resulting conductance is determined by
\textit{individual} hopping processes, and not by an average over
such processes (as required for the bulk). Indeed, in this case
the tunnelling trajectory for a ``critical hop"  passes through
the small contact region, see Fig.~\ref{fig:fig1}, and, in
general, it is not straight. As a result, a typical trajectory
connecting  the two sides of the point contact, which determines
the conductance of the whole system, is longer than the typical
hop in the bulk.

In this paper we show that the {\it magnetoresistance} of such a
point contact provides further information about the properties of
the individual localized states at its vicinity. The contact can
also serve as a detector for the local spin polarization. We
discuss several mesoscopic effects revealed by the application of
the magnetic field: (a) the shrinkage of the wave functions, (b)
Aharonov-Bohm oscillations, (c) spin accumulation and correlations
near the contact, and (d) the Hall effect.

In addition to discussing such individual hops, we also discuss
the statistics of these hops, when the experiment is repeated over
many realizations of the sample or of the Fermi level within one
sample. The statistical properties of the resistance and of the
magnetoresistance of point contacts are found to reflect the
spatial and energy distributions of the individual hopping states
near the contact. The calculated distribution function of the
magnetoresistances at small values follows a power law; at large
values it decays as a stretched exponential.   The Aharonov-Bohm
oscillations associated with the contact allow the determination
of the  occupation numbers of the relevant states, as well as the
strength of the impurity potential. Spin-dependent effects reveal
spin correlations and the point contact can thus serve as a
detector of the local spin accumulation or depletion. Finally, the
mesoscopic Hall effect is sensitive to the correlation length in
the bulk, since the percolation cluster shunts the effective
``Hall generator'' located near the contact. Thus, the point
contact spectroscopy is a powerful tool for \textit{quantitative
studies} of the nature and of the parameters of individual
localized states.

There are two reasons for the large difference between the
conductance of the bulk and the conductance through a point
contact.  First, as shown in Ref. \onlinecite{KZ}, the total
length of a typical ``critical hop" between the sites located near
the contact in different half-planes (or half-spaces) is about
twice larger than for hopping in the bulk. Correspondingly, the
resistance associated with this ``critical hop" is exponentially
larger than that associated with a typical hop in the bulk. This
leads to specific statistical properties of the contact resistance
and magnetoresistance. The second reason is related to the
different geometries of the percolation cluster in the bulk, when
the point contact is present or absent.

Let us start with the second effect. The necessity to go via the
contact yields an additional constraint on the choice of the hopping
sites, compared with the choice for the bulk conductor. To estimate
the expected increase of the resistance, we consider first the bulk
resistance, $R=R_0e^\xi$ (this is the definition of $\xi$). For the
case of Mott hopping, this resistance is determined by the effective
resistor network percolation cluster, for which $\xi$ exceeds its
critical percolation threshold, $\xi_c=(T_0/T)^{1/(d+1)}~(\gg 1)$ in
$d$ dimensions (here, $T_0 \propto 1/[g a^d]$, where $g$ is the
density of states at the Fermi energy, and $a$ is the decay length of
the localized states).\cite{Shklovskii}  Taking $\xi$ within the
interval
\begin{align}
\xi_c < \xi < \xi_c +\Delta \xi \ ,
\end{align}
the typical correlation length of the percolation cluster can be
estimated by \cite{Shklovskii}
\begin{align} {\cal L}_{\Delta \xi}
\simeq r_h [\Delta \xi/\xi_c]^{-\nu}\ ,
\end{align}
where $r_h \approx a \xi_c/2$ is the hopping length (which
replaces the usual lattice constant for the percolation problem)
and $\nu$ is the percolation correlation length exponent (equal to
4/3 in two dimensions). For a bulk HI, $\Delta \xi \approx 1$ and
the correlation length is ${\cal L} \sim \xi_c^{\nu} r_h$. In the
present case, which pertains to two dimensions, the actual
percolating cluster can then be thought of as being built of a
regular two dimensional structure, made of random blocks of size
${\cal L}$. The final resistance $R$ is then obtained by averaging
the log of the resistance over these blocks.

In the presence of the point contact, the percolating cluster
\textit{must} pass via the orifice, with a probability of the
order of unity. A sufficient condition for this is
\begin{align}
r_h < {\cal L}_{\Delta \xi} \leq D\ ,\label{bound}
 \end{align}
namely $\Delta \xi \geq \xi_c (r_h/D)^{1/\nu}$. Otherwise, the
orifice may be surrounded by a finite cluster, which is
disconnected from the bulk. This increase in $\xi$ implies an
increase in the total resistance, compared to that of the bulk HI.
Thus, the resistance of the system may be larger than $R_0
e^{\xi_c[1+(r_h/D)^{1/\nu}]}$. Given Eq. (\ref{bound}), this may
be as large as $R^{}_0 e^{2\xi_c}$, but the temperature dependence
of the resistance is still dominated by that of $\xi_c$, which is
similar to that of the bulk (although it is renormalized).

In this paper we concentrate on the case $r_h \gg D$. In that
case, we argue that the total resistance is dominated by that of
the individual hop near the orifice. As we show below, this
situation generates a different temperature dependence of the
resistance, which allows one to discriminate between the two
situations experimentally.

The outline of the paper is as follows: Section II discusses the
magnetoresistance, and shows how measurements of both the
resistance and the magnetoresistance allow the extraction of
detailed information on the individual localized states which
dominate the hopping. This section also discusses the distribution
of the magnetoresistance. Section III discusses the Aharonov-Bohm
oscillations, due to the quantum interference between close paths
near the point contact. In Sec. IV we show that the amplitude of
these oscillations depends on the site occupation and on the spin
configuration of the relevant localized states, and argue that
this dependence can be used to identify spin accumulation.
Finally, Sec. V discusses the mesoscopic Hall effect due to the
individual configuration near the point contact.

\section{Magnetoresistance}

From now on we restrict the discussion to two-dimensional devices,
and model the point contact by an orifice in the thin but
infinitely strong barrier dividing the two-dimensional plane into
the two semi-infinite half-planes, see Fig.~\ref{fig:fig1}.
Similar arguments also work in three dimensions. The orifice
diameter, $D$, is taken such that $a \ll D \ll n^{-1/2}$ where $n$
is the concentration of the hopping centers. We next concentrate
on one critical hop, between two sites on the two sides of the
orifice, with coordinates  ${\bf r}_1$ and  ${\bf r}_2$ (both
measured from the bottom of the orifice). The assumption $D \ll
n^{-1/2}$ ensures that there will be only one such site on each
side of the orifice. A relatively weak magnetic field
$\mathbf{H}$, such that $\lambda=(c\hbar/eH)^{1/2} \gg a $, is
applied normally to the plane.

We assume that the thickness of the sample, $t$,  satisfies the
conditions
\begin{align}
  \label{eq:01}
  a \ll t \ll {\cal L}\, .
\end{align}
The inequality $t \ll {\cal L}$ ensures the two dimensional
character of hopping, and the inequality $t \gg a$ allows using
the three dimensional asymptotic behavior of the localized wave
function. Our main results also hold for $D \gtrsim a$.

The wave function of a hydrogen-like isolated localized state in a
magnetic field can be written as (see, e.~g.,~Ref.~\onlinecite{Shklovskii}),
\begin{eqnarray}\label{psi}
\Psi(\br,{\bf r}_i)=\frac{1}{\sqrt{\cal N}}\,e^{-A(\br,\br_i)},
\quad {\cal
 N}=\int d^3 r \, e^{-2 A(\br,0)}\, ,
 \nonumber \\
A(\br,\br_i)=\frac{|\br -\br_i|}{a}+\frac{|\br -\br_i|^3a}{24\lambda^4}
 - \frac{ie}{2 \hbar c}\left[ \bH \times \br_i \right]\cdot \mathbf{r}\ .
\end{eqnarray}
The wave function shrinkage, described by the factor
$e^{-r^3a/24\lambda^4}$, originates from the centrifugal potential
induced by the magnetic field.

We next discuss the overlap integral, $V_{12}$,  between the wave
functions centered at $\br_1$ and at $\br_2$. As long as we do not
consider interference between this hop and other hops, the last
term in $A(\br,\br_i)$ only represents a phase factor, which does
not affect the magnitude of the overlap between the wave functions
at sites 1 and 2. Ignoring these phase factors, one has
\begin{align}\label{pm1}
V_{12}\sim V_0 e^{-(r_{1}+r_{2})/a} \, e^{-a\left( {r_1^3} +
{r_2^3} \right)/24\lambda^4}, \ V_0 \sim e^2/\kappa a,
\end{align}
where $\kappa$ is the dielectric constant. The magnetoconductance
ratio $G(H)/G(0)$ is then given by the expression
\begin{equation}
  \label{eq:gh}
  G(H)/G(0) =e^{- a(r_1^3+r_2^3)/12\lambda^4}\ .
\end{equation}

At low fields, it is reasonable to assume that the same pair of
centers, at distances $r_1$ and $r_2$, dominate both the zero
field resistance and the low field magnetoresistance.
Interestingly, measurements of both the resistance and the
magnetoresistance can yield information on the individual hop near
the orifice. For the given pair of sites, the zero-field
conductance is given as
\begin{align}\label{cond}
G_{12}=G_0\exp \left[- 2 (r_1 + r_2)/a - \Delta \varepsilon/T
\right]\ ,\end{align}
 where $\Delta \varepsilon$ is the activation
energy. This conductance is definitely distinct from the bulk
variable range hopping one, which is given by the Mott law,
\begin{align} G=G_0 e^{-\xi_c}, \quad \xi_c=(T_0/T)^{1/(d+1)}.
\end{align}
In fact, Eq. (\ref{cond}) exhibits an Arrhenius activation temperature
dependence. Observing such a temperature dependence can indeed confirm
that the conductance is dominated by the individual hop near the
orifice, and not by the whole percolation cluster, as discussed at the
end of the previous section.  Measuring this temperature dependence
can then yield estimates for both $\Delta \varepsilon$ and
$G_0e^{-2(r_1+r_2)/a}$. Measuring the conductance of a similar sample,
without the orifice, would yield estimates for $T_0$ and for
$G_0$. Assuming that $G_0$ is roughly the same in both cases (the
difference in $G_0$ only adds a logarithmic correction to the
exponent), one can end up with an estimate of the combination ${\cal
A}=(r_1/a)+(r_2/a)$.

As stated in Ref. \onlinecite{Shklovskii}, the magnetoconductance
of a similar bulk system has the form
\begin{align}
\frac{G(H)}{G(0)}=\exp \left[-t_1 \left (\frac{a}{\lambda}\right
)^4\left (\frac{T_0}{T}\right )^{d/(d+1)}\right ]\ ,
\end{align}
where $t_1$ is a numerical factor. Since we know $T_0$ from the
conductance, and we know the $H$-dependence of $\lambda$, we can
use this information for estimating $(a/\lambda)$. Measuring Eq.
(\ref{eq:gh}) can thus yield estimates for the combination ${\cal
B}=(r_1/a)^3+(r_2/a)^3$. Combining these two equations, one finds
$R_{1,2}/a={\cal A}/2\pm \sqrt{{\cal B}/(3{\cal A})-{\cal
A}^2/12}$.

It should be noted that the choice of the two sites which dominate
the conductance may vary with temperature.\cite{KZ} In fact, one
may end up with different Arrhenius curves for different
temperature ranges. The above method can thus yield details about
the individual pairs of sites which dominate the hopping in each
such range.  In principle, this information can be used for
building up the distribution of the hopping sites near the orifice
in real space and in energy.

Interestingly, the magnetoresistance  allows one to find whether
the transport is dominated by the states close to the contact. For
bulk measurements, the conductance yields the hopping length $r_h$
and then $\log[G(H)/G(0)]\propto r_h^3$. In contrast, when the
conductance is dominated by the contact then $r_h$ is replaced by
 $r_{12} =r_1 +r_2$. We next assume that for the most probable configuration
one has $r_1 \sim r_2 \sim \bar{r}\equiv(\pi n)^{-1/2}$, where $n$
is an effective concentration of the hopping sites relevant for
the conductance (In the cases of the Mott or the Efros-Shklovskii
conductance, this concentration is temperature-dependent). With
this assumption, one has $\log[G(h)/G(0)]\propto r_1^3+r_2^3\sim
2{\bar r}^3 \sim r_{12}^3/4$, namely about 4 times smaller than it
would be expected for the bulk magnetoresistance for the same
hopping length.

\begin{figure}[h]
\centerline{ \includegraphics[width=5cm]{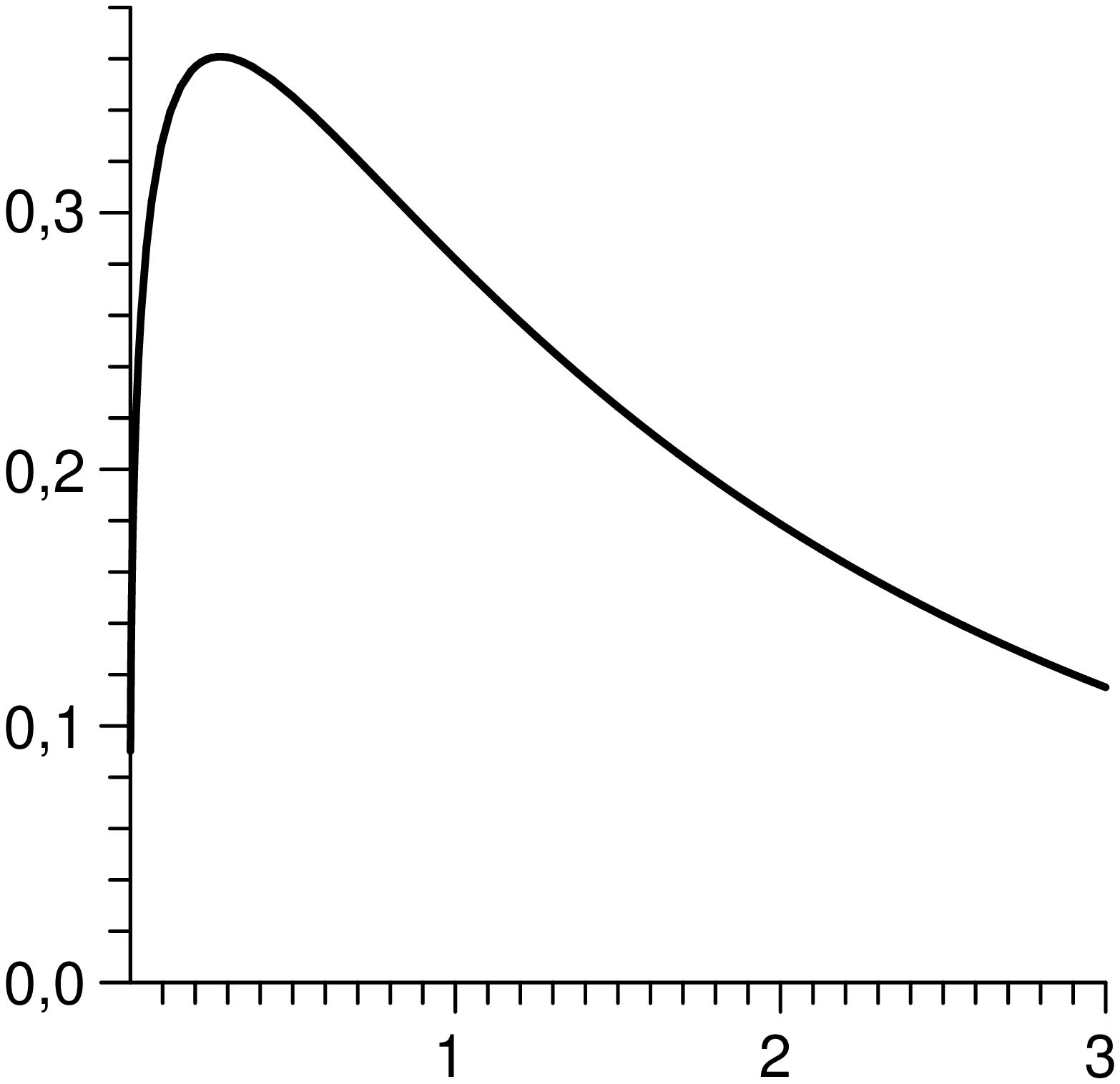}}
\caption{Graph of the function $\cF(z)$, Eq. (\ref{eq:df02}).}
\label{fig:df}
\end{figure}

We next discuss the statistics of the magnetoresistance, when the
measurements are done on many realizations of similarly prepared
samples. For this purpose, we introduce the relative
magnetoresistance, $\rho \equiv 1-G(H)/G(0)$. We also use the
typical value for the exponent in Eq.~(\ref{eq:gh}),  $r_1
=r_2=\bar{r}=(\pi n)^{-1/2}$.  The characteristic
magnetoresistance is then
 \begin{equation}
   \label{eq:barg}
   \bar{\rho}(H) \equiv a\bar{r}^3/6\lambda^4 \ll 1\, .
 \end{equation}
Here we have taken into account that $\lambda \gg \bar{r}$. Equation
(\ref{eq:gh}) then implies that
\begin{equation}
  \label{eq:mr1}
  \rho=\bar{\rho}(r_1^3+r_2^3)/2\bar{r}^3\, ,
\end{equation}
and the quantity $\bar{\rho}$ can be treated as a typical value for
the magnetoresistance. We next recall our assumption of low
concentration, $D \ll n^{-1/2}$, which implies that there is only one
relevant scatterer on each side of the orifice. At low fields, the
same two centers will dominate both the zero field resistance and the
magnetoresistance. Under this assumption, the two distances $r_1$ and
$r_2$ are independent of each other, and each of them is characterized
by a Poisson distribution, $\cP(r)=(2/
\bar{r}^2)\,e^{-(r/\bar{r})^2}$. The distribution of the $1 \to 2$
magnetoresistance then becomes (see Appendix \ref{DERIV} for details)
\begin{align}
\cP(\rho)=\int \cP(r_1)dr_1 \int \cP(r_2)dr_2 \, \delta [\rho - \rho
(r_1, r_2)].
\end{align}
This distribution then obeys the scaling form
$\cP(\rho)=(1/\bar{\rho})\cF(\rho/\bar{\rho})$, where
\begin{equation}
  \label{eq:df02}
  \cF(z)=\frac{4z^{1/3}}{9}\int_0^1 \frac{d\xi\,e^{-z^{2/3}[\xi^{2/3}+
(1-\xi)^{2/3}]} }{\xi^{1/3} (1-\xi)^{1/3} }\, ,
\end{equation}
see the plot in Fig.~\ref{fig:df}. This scaling function has the
limiting behaviors $\cF(z) \propto \sqrt{z}$ at small $z$, and $\cF(z)
\propto e^{-z^{2/3}}$ for large $z$. Thus at small enough $r_i$ the
magnetoresistance has only a power law dependence on $r_i$, in
contrast to the contact resistance itself. Needles to say, this limit
represents very rare realizations.

At very strong magnetic fields, where the exponent
$a\bar{r}^3/6\lambda^4$ becomes comparable to $\xi_c$, the contact
resistance can be lowered by including hopping between a pair of sites
with site energies outside of the hopping energy band, since the
concentration of the corresponding sites is larger and thus their
typical spatial separation is smaller. This additional hopping reduces
the magnetoresistance. Such ``switching" between the different pairs
with an increasing  magnetic field is expected to lead to giant
mesoscopic fluctuations of the magnetoresistance, which are similar to
the fluctuations of the conductance as a function of temperature and
applied bias, considered in Ref.~\onlinecite{KZ}.

\section{Aharonov-Bohm effect}

\begin{figure}[h]
\centerline{ \includegraphics[height=4cm]{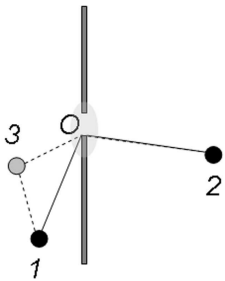}}
\caption{The  triangle which determines the interference effects
near the orifice.} \label{fig3}
\end{figure}

The presence of a third localized state, at $\br_3$ near the point
contact, yields a sample specific Aharonov-Bohm effect, i.~e.,
oscillations of the magnetoconductance due to interference between
different tunnelling paths.  We next evaluate the interference between
the ``direct" tunnelling trajectory, $1 \to 2$, which hits the
orifice, and the ``scattered" trajectory, that starts at the same
initial center, $1$, and reaches the final center, $2$, after being
scattered by the scatterer $3$. A typical arrangement of the centers
is shown in Fig.~\ref{fig3}.
The trajectories interfere on the left side of the orifice, and almost
coincide in the opposite half-space. The interference triangle for the
case shown in Fig.~\ref{fig3} is  $1-3-O$, the interfering paths being
$1 \to O$ and $1 \to 3 \to O$. We next follow the  perturbative
approach used by Raikh and Wessels.\cite{Raikh} To leading order, the
terms which involve site 3 contain denominators like $(\varepsilon_3 -
\varepsilon_1)$ and $(\varepsilon_3 - \varepsilon_2)$. The energy
$\varepsilon_3$ is spread within a band with a width of order
$e^2/\kappa a \gg \Delta$, where $\Delta$ is the width of the hopping
band. Since the energies $\varepsilon_1$ and $\varepsilon_2$ belong to
the hopping band, typically one has $|\varepsilon_3| \gg
|\varepsilon_{1,2}|$. Therefore, we approximate these denominators by
$\varepsilon_3$. The contribution of the triangle to the conductance
then becomes
\begin{eqnarray} \label{AB}
&& G \propto
\frac{|V_{12}\left(1+Je^{i\varphi}\right)|^2}{(\varepsilon_1 -
\varepsilon_2)^2} =\frac{|V_{12}|^2(1 + J^2 + 2J\cos
\varphi)}{(\varepsilon_1 - \varepsilon_2)^2}; \nonumber \\
&&V_{ij}=V_0e^{-r_{ij}/a}, \ J \approx
\frac{V_{13}V_{32}}{V_{12}\varepsilon_3}, \ \varphi = 2 \pi
\frac{HS}{\Phi_0}\, .
\end{eqnarray}
Here $r_{ij}$ is the distance between the sites $i$ and $j$
\textit{along the tunnelling trajectory}, $S$ is the area of the
interference triangle $1-3-O$ and $\Phi_0= 2\pi \hbar c/e$ is  the
flux quantum.

The typical area of the triangle, $\bar{S}$, can be estimated from
the following considerations, cf. Ref.~\onlinecite{Raikh}.
Consider the triangle $1-3-O$ in Fig.~\ref{fig3}. For a symmetric
configuration, the difference between the lengths $1-3-O$ and
$1-O$ is $2h^2/R$ where $h$ is the height of the triangle while
$R$ is the distance between 1 and $O$. As in
Ref.~\onlinecite{Raikh}, we restrict ourselves to configurations
where this difference does not exceed $a$, so that typically $h
\sim ({\bar R}a)^{1/2}$ and $\bar{S} \sim \bar{r}^{3/2}a^{1/2}$.
Otherwise, the ratio $V_{13}V_{32}/V_{12} \sim
V_0e^{-(r_{13}+r_{3O}-r_{1O})/a}$ becomes exponentially small, and
the oscillations will not be visible. The probability for this
restriction to hold is of the order of the ratio between $\bar{S}$
and the typical area per impurity, ${\bar r}^2$, i.e.
$\sqrt{a/\bar{r}}$. Since most experiments are done close to the
metal-insulator transition, this probability need not be very low.
Once this restriction holds, then we may encounter large values of
$|J|$ in Eq. (\ref{AB}), even for $\max ( |V_{13}|, |V_{32}|,
|V_{12}|) << 1$. The sign of $J$ can be arbitrary, depending on
whether the energy of state 3 is above or below the Fermi energy.

The characteristic field for the Aharonov-Bohm effect, $H_c
=\Phi_0/\bar{S} \sim \Phi_0/(\bar{r}^{3/2}a^{1/2})$, is of the same
order as the critical field for the positive magnetoresistance
originating from the shrinkage of the wave functions.  Thus the
Aharonov-Bohm oscillations decay strongly for $H>H_c$.  However, one
can expect that an oscillation pattern will be observed in addition to
the smooth increase of the resistance with the magnetic field, since
the relative oscillation amplitude, $2J/(1+J^2)$ can be of the order
of unity.

For Coulomb scattering centers with $V_0 \sim e^2/\kappa a \gtrsim
\Delta$, we can follow the estimates of Ref.  \onlinecite{Raikh} and
conclude that the quantity $J$ is of order 1 (``strong
scattering"). In the opposite case of weak scattering, $V_0 \sim
e^2/\kappa r$, $r$ being a distance from the scattering center, the
dependence of the matrix element on the hopping distance becomes
important.~\cite{Agrin} Since in this case $J \ll 1$, the amplitude of
the Aharonov-Bohm oscillations is significantly less than that for the
case of strong scattering.  Correspondingly, the interference
contribution decreases with the temperature  due to the growth of the
hopping length.

\section{Non-equilibrium spin accumulation}

The Aharonov-Bohm effect in a point contact can be used to
\text{detect non-equilibrium spin accumulations} in hopping
systems. Following Ref.~\onlinecite{Spivak}, we consider the situation
where  site 2 is empty and  site 1 is occupied. If site 3 is also
occupied, the tunnelling processes can be considered as tunnelling of
the \textit{hole} from  site 2 to site 1. The interference takes place
only if the electron spins on sites 1 and 3 are parallel. Indeed, the
final configurations for the paths $2 \to 1$ and $2 \to 3 \to 1$ are
the same only for parallel 1-3 spins. Consequently, the overlap
integral for this case is given by Eq.~(\ref{AB}). Otherwise, the
interference contribution is absent. When both sites 2 and 3 are empty
(tunnelling of an electron from 1 to 2), the interference effect is
also present.

{}From Eq.~(\ref{AB}) we derive the modulation depth of the
conductance oscillations as
\begin{eqnarray}
 % \label{eq:md}
 \cM  \equiv  \frac{\cG_{\max} - \cG_{\min}}{\cG_{\max} + \cG_{\min}}
 = \frac{2|J|}{1+J^2} \left[n_{1\ua}n_{3\ua} +n_{1\da}n_{3\da}
 \right.\nonumber \\    \left.  +n_{1}(1-n_{3} ) \right](1-n_{2})
 =\frac{|J|}{1+J^2} (4s_1s_3n_3 - n_3 +2 ).\label{16}
\end{eqnarray}
Here $n_{i\sigma}$ is the occupation number of  spin $\sigma$ at
site $i$, $n_i \equiv n_{i\ua}+n_{i\da}$ is the total occupation
of the site $i$ (which is limited to be 0 or 1, due to strong
on-site interactions), while $s_i \equiv (n_{i\ua}-n_{i\da})/2$ is
its spin accumulation.  In deriving the last equality in Eq.
(\ref{16}), we have assumed that $n_1=1,n_2=0$, corresponding to
the case of low temperatures and to hopping from left to right.
The first term in the brackets is sensitive to the site spins.
Indeed, if there is no spin polarization, and if there are also no
spin correlations, then on average $\langle s_1s_3\rangle =\langle
s_1 \rangle \langle s_3\rangle= 0$.  If there is a non-equilibrium
source of spin accumulation or depletion, then the modulation
depth will change, signaling the related local polarization.

We next apply an external magnetic field, in equilibrium. Writing the
Zeeman energy as $-g\mu^{}_B H s$, where $\mu^{}_B=e\hbar/(2mc)$ is
the Bohr magneton and $g$ is the $g$-factor, we would have $\langle
s_i \rangle=\frac{1}{2}\tanh (H/H_p)$, where $H_p= 2kT/(\mu^{}_B
g)$. If ${\bf H}$ is in the plane of the triangle, then $\cos \varphi
=1$, and the magnetoresistance would become
\begin{equation}
\frac{\delta G}{G} =  \frac{J}{1 + J^2}n_3\tanh^2 (H/H_p).
\end{equation}

When the field is perpendicular to the plane, then the
magnetoresistance has both the Aharonov-Bohm oscillations, at a
typical scale $H_c$, and the above contributions, which saturate for
$H \gg H_p$. The ratio of these scales is given by
\begin{align}\frac{H_p}{H_c}= \frac{kT}{\pi
g(\hbar^2/ma^2)}\left(\frac{\bar{r}}{a}\right)^{3/2}.
\end{align}
Since $\hbar^2/ma^2$ is of the order of the Bohr energy, say 200K, and
since experiments would be typically done at a few degrees K, this
ratio would become of order unity only when $\bar{r}/a \sim 30$, but
then it would be difficult to observe any conductance.  Thus, usually
$H_p \ll H_c$, and the spins would be saturated in the amplitude of
the Aharonov-Bohm oscillations.

Since the field $H_p$ is well-defined, studies of the
magnetoresistance in relatively weak magnetic fields can help
calibrating the device, i.e., finding $J$ and $n_3$. Indeed, if $n_3 =
1$, the system exhibits a positive magnetoresistance, which saturates
at $H > H_p$.  Thus one should look for structures in which the
positive magnetoresistance mentioned above is observed (which means
that $n_3 = 1$) and then obtain $J$ from the saturated value of this
magnetoresistance according to the formulae given above. Here we may
profit from the fact that the saturation field depends only on
well-defined values of $T$ and $g$, and is thus easily specified.

Note that the spins can be aligned by a magnetic field in {\it any}
direction, while the Aharonov-Bohm effect is sensitive only to
magnetic fields perpendicular to the plane of the tunnelling.  Thus,
even in the case when $H_p \sim H_c$ one can separate spin effects by
applying the magnetic field parallel to the tunnelling plane. The
effect depends on the product $s_1s_3$, and therefore it is
independent of the \textit{sign} of the spin polarization.  This
offers an alternative way for the observation of the spin accumulation
which is generated by an ac electric field, as predicted in
Ref.~\onlinecite{Entin}.

\section{Mesoscopic Hall effect}

In this paper we concentrate on the situation when the hopping
through the point contact is dominated by the vicinity of this
contact. In this case, the bias voltage is completely concentrated
on the ``critical" pair of sites (1 and 2) near the orifice. As a
result, we expect the Hall voltage also to be generated only by
the single triangle of sites 1--3--O  (Fig.~\ref{fig3}), and not
to be affected by any averaging procedure. This triangle then
serves as an elemental ``Hall generator''. In the presence of an
external magnetic field, $H$,  perpendicular to the sample, this
triangle generates an additional contribution to  the potential
difference between the sites 1 and 3 when the charge transfer
occurs between the sites 1 and O. Consequently, a Hall voltage,
$V_H$, is generated between the (transverse) sample edges.

\begin{figure}[h]
\centerline{\includegraphics[width=\columnwidth]{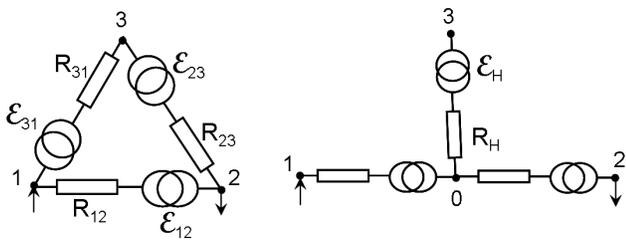} }
\caption{Equivalent circuit of a Hall generator. The resistances
and electromotive forces are calculated from the Miller-Abrahams
resistances between the sites and the additional  voltage
differences induced by the magnetic field.}\label{fig:star}
\end{figure}

One can present the triad 1--2--3 as the equivalent circuit shown
in the left panel of Fig.~\ref{fig:star}. Here $R_{ik} \equiv
G_{ik}^{-1}$ are the Miller-Abrahams\cite{abrahams} resistors
found from Eq.~(\ref{cond}).
The magnetic field changes the hopping probabilities of the
inter-site hopping. As shown in Ref.~\onlinecite{Karpov}, the
changes of the potentials on the sites can be taken into account
by introducing effective electromotive forces (EMFs), $\cE_{ik}$,
proportional to the current entering through the site 1 and
leaving through the site 2.

These EMFs vanish when $H=0$. When $H\ne 0$, they can be
expressed through the changes of the hopping probabilities in a
non-zero magnetic field,\cite{Holstein} and they turn out to be
proportional to $H$. In order to estimate the voltage between the
Hall leads it is convenient to transform the equivalent circuit to
the ``star'' configuration shown in the right panel of
Fig.~\ref{fig:star}. The quantity $\cE_H$ then represents the EMF
of the Hall generator, while $R_H$ is its internal resistance. The
approximate expression for $\cE_H$ at $|\varepsilon_i| \gtrsim T$
is~\cite{Karpov}
\begin{eqnarray}
  \label{eq:emf1}
  \cE_H&=&V_{12}\frac{\bH \cdot
  \bS}{\Phi_0}
  \frac{R_{31}R_{23}(R_{12}+R_{23})}{R_{12}(R_{12}+R_{23}+R_{31})^2}(1-n_3)
  \beta \, , \nonumber \\
\beta&=& \text{sign} (V_{12}V_{23}V_{31})\, \frac{\hbar }{e^2R_0}\frac{2\pi
  TV_0}{(\varepsilon_1-\varepsilon_2)^2}\,
  e^{(r_{12}-r_{13}-r_{23})/a}
\nonumber \\ &&\times
\frac{e^{-\varepsilon_1/T}+e^{-\varepsilon_2/T}}{e^{-\varepsilon_1/T}+
  e^{-\varepsilon_2/T}+e^{-\varepsilon_3/T}}\, , \nonumber \\
R_H&=&\frac{R_{31}R_{23}}{R_{12} + R_{23}+R_{31}}\, .
\end{eqnarray}
Note that the variation of the hopping probability by the magnetic
field is determined by two-phonon processes,~\cite{Holstein} and
consequently $\beta \ll 1$. The typical internal resistance,
$R_H$,  of the elemental Hall generator is of the the order of the
hopping resistor, $R_h$.

\begin{figure}[h]
\centerline{
\includegraphics[width=6cm]{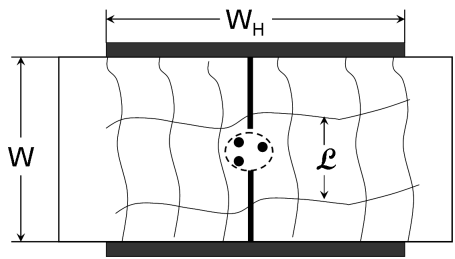}
}
\caption{Illustration of the connection between the Hall generator
  and the leads. \label{fig:Hall}}
\end{figure}

Now let us estimate the voltage between the Hall leads, see
Fig.~\ref{fig:Hall}.
The generator is connected to the Hall leads by a branch of the
percolation cluster. This chain is bypassed by the resistance between
the Hall electrodes, which consists of $W_H/\mathcal{L}$
branches. Here $W_H$ is the width of the Hall electrodes while $\cL$
is the correlation length of the percolation cluster. This ratio must
be large, otherwise the Hall voltage would not be measurable because
it would be subjected to very strong fluctuations. Thus the measured
Hall voltage is smaller than the Hall voltage of a single triad, by
the factor $\cL/W_H \ll 1$,
\begin{equation}
    \label{eq:hall1}
    \frac{V_H}{V} \approx \beta \frac{HS}{\Phi_0}\, \frac{\cL}{W_H}\, .
  \end{equation}
One can expect that the mesoscopic Hall effect will exceed the bulk
contribution of the electrodes. Assuming that almost all the voltage
$V$ drops at the contact region and only its small part, $\eta V$
drops at the leads, and using the estimate for the bulk contribution
to the Hall voltage from Ref.~\onlinecite{Karpov}, we obtain the
order-of-magnitude estimate
\begin{equation} \label{eq:hg}
\frac{V_H}{V_{H}^\text{(bulk)}}\approx \frac{\mathcal{L}
  \mathcal{L}_H}{\eta Wr_h}  \approx \frac{\xi}{\eta}\,
  \frac{\cL_H}{W}\, .
\end{equation}
Here we have assumed that $W \approx W_H$, while $\cL_H \gg \cL$ is
the correlation length of the network of Hall generators in the
bulk.\cite{Karpov} This ratio can be large due to the exponential
smallness of $\eta$; this factor can be independently estimated as the
ratio of the resistance in the bulk and the contact resistance.

\section{Summary}

Let us finally summarize the procedure for extracting local parameters
of the point contact and the adjacent states. Studying the
Aharonov-Bohm oscillations, one can extract the occupation at site 3,
$n_3$,  and the quantity $J$, Eq. (\ref{AB}), from the amplitude of
these oscillations. Similar information, as shown above, can be
obtained from the weak-field positive magnetoresistance resulting from
the spin alignment. Since the presence of these effects ensures that
$n_3 = 1$, one can also find the average $\langle s_1 s_3 \rangle$,
which signals the presence of non-equilibrium spin accumulation or
depletion.  Indeed, at equilibrium one has $\langle s_1s_3\rangle =
0$, while a complete spin alignment gives $\langle s_1s_3\rangle =
1/4$. In this way, the point contact can be ``calibrated''.

To conclude, we have considered the influence of an external
magnetic field on the mesoscopic conductance of a point contact
between two hopping insulators. We have shown that the applied
magnetic field allows the identification and the measurement of
spin effects. We discussed the mesoscopic Hall effect and showed
that it provides important information about the hopping cluster
correlation length. All of these effects provide information about
hopping processes associated with a \textit{single} hopping event,
not distorted by an averaging procedure. Furthermore, the
statistics of the hopping transport through a point contact can,
in principle, provide information about the distribution of the
hopping states near the contact. In gated structures such
statistics can be studied by changing the gate voltage in a
\textit{single} device.

\appendix
\section{Derivation of the distribution function}
\label{DERIV}

 We have $\rho= \rho_1+\rho_2$, where
$\rho_i=\bar{\rho} (r_i/\bar{r})^3$. Then
\begin{eqnarray}
  \label{eq:ap1}
  \cP(\rho)&=&4\int d\rho_1 d\rho_2\,  \delta(\rho_1+\rho_2-\rho)
 \nonumber \\&& \times \prod_{i=1,2}
 \int x_i\, dx_i
 e^{-x_i^3}\delta\left(\rho_i-\bar{\rho}x_i^3 \right)\,
\end{eqnarray}
Since $\int x\, dx \delta\left(\rho -\bar{\rho}x^3
\right)=(3\bar{\rho})^{-1} \left(\bar{\rho}/\rho \right)^{1/3}$,
we readily obtain Eq.~(\ref{eq:df02}).

\acknowledgments

 This work was supported by the U. S. Department of
Energy Office of Science through contract No.DE-AC02-06CH11357, by a
Center of Excellence of the Israel Science Foundation 
(at BGU), and by a grant from by the German Federal Ministry of
Education and Research (BMBF) within the framework of the
German-Israeli Project Cooperation (DIP) (at BGU).

\end{document}